\documentstyle[galley, rotate]{mn}
\onecolumn
\input{epsf}

\def\inc{{\int_0^{\chi_s}}}

\title[Weak Gravitational Lensing]
{Weak Lensing from Strong Clustering}

\author[D. Munshi and  P. Coles]{Dipak Munshi$^{1}$ and Peter Coles$^{2}$\\
 $^1$Max-Planck-Institut fur Astrophysik,
Karl-Schwarzschild-Str.1, D-85740, Garching, Germany
\\$^2$School of Physics \& Astronomy, University
of Nottingham, University Park, Nottingham, NG7 2RD, United
Kingdom\\}

\pagerange{000,000}

\begin{document}
\maketitle

\begin{abstract}
We investigate the effect of weak gravitational lensing in the
limit of  small angular scales where projected galaxy clustering
is strongly nonlinear. This is the regime likely to be probed by
future weak lensing surveys. We use well-motivated hierarchical
scaling arguments and the plane-parallel approximation to study
multi-point statistical properties of the convergence field. These
statistics can be used to compute the vertex amplitudes in tree
models of hierarchical clustering; these can be compared with
similar measurements from galaxy surveys, leading to a powerful
probe of  galaxy bias.
\end{abstract}


\section{Introduction}

Weak gravitational lensing by large-scale density perturbations is
likely to be one of the most useful tools for studying the
statistical properties of the distribution of gravitating matter
and the geometry of the background universe (Blandford et al.
1991; Miralda-Escud\'{e} 1991; Kaiser 1992). Studies of lensing in
the weak limit usually revolve around the statistics of the {\em
convergence} field $\kappa$, which is related to the line-of-sight
fluctuations in gravitational potential and thus to the
distribution of gravitating mass; for recent technical reviews see
Kaiser (1996), Bernardeau (1999) and Mellier (1999).

Various methods, both analytical and numerical, have been applied
to understand weak lensing. Bernardeau et al. (1997) have applied
gravitational perturbation theory to compute statistical
properties of $\kappa$. Numerical computations have mainly
focussed on ray-tracing experiments using N-body simulations.
Examples of this approach include: Jain \& Seljak (1997); Jain,
Seljak \& White (1998, 1999a, 1999b); Premade, Martel \& Matzner
(1998);  van Waerbeke, Bernardeau \& Mellier (1998);  Bartelmann
et al. (1998); and Couchman, Barber \& Thomas (1998). This method
has the advantage that it can treat the small angular scales where
perturbative calculations are no longer applicable. It is,
however, difficult to probe a large parameter space of possible
models using numerical experiments and it is consequently
important to investigate analytic methods for studying the regime
where non-linear gravitational fluctuations influence the lensing
of light rays.

In this paper we use a hierarchical model, or {\em ansatz}, for
the form of the $n$-point correlation functions of matter in the
non-linear regime. Such models have a long history (Groth \&
Peebles 1977; Davis \& Peebles 1977; Fry \& Peebles 1978;
Bernardeau \& Schaeffer 1992; Szapudi \& Szalay 1993) but have
recently been shown to generate
simple and accurate predictions for  count-in-cell statistics
(Munshi, Coles \& Melott 1999a,b), the mass function of collapsed
objects (Valageas \& Schaeffer 1999) and  the bias and
higher-order correlations of over-dense cells (Bernardeau \&
Schaeffer 1992; Munshi et al. 1999a; Bernardeau \& Schaeffer 1999;
Coles et al. 1999). Numerical simulations have also demonstrated
that the hierarchical ansatz is very accurate in predicting
clustering properties (Munshi et al. 1999b). It is natural to use
this model to study the statistical properties of the convergence
field involved in weak lensing calculations.

\section{Matter Correlations and Convergence Statistics}

The statistics of $\kappa$ are very  similar to those of the
projected density field. The formalism for handling higher-order
correlation functions of projected galaxy clustering was developed
by T\'{o}th, Holl\'{o}si \& Szalay (1989) and later applied by,
e.g., Gazta\~{n}aga (1994) and Frieman \& Gazta\~{n}aga (1994).
The relevance and usefulness of these methods to studies of the
convergence has been pointed out by Gazta\~{n}aga \& Bernardeau
(1998) and Hui (1999). The results we present here are new because
they concern a different kind of statistical descriptor, the
cumulant correlator. We begin this paper, therefore, by developing
the basic relationship between cumulants (and cumulant
correlators) of the convergence field $\kappa$ with the underlying
statistics of the matter distribution.

We adopt the following line element for background geometry:
\begin{equation}
d\tau^2 = -c^2 dt^2 + a^2(t)[ d\chi^2 + r^2(\chi)d^2\Omega],
\end{equation}
where $a(t)$ is the expansion factor. The angular diameter
distance $r(\chi)= K^{-1/2}\sin [K^{-1/2} \chi]$ for positive
spatial curvature, $r(\chi) = (-K)^{-1/2}\sinh [(-K)^{-1/2}\chi] $
for negative curvature, and $r(\chi)=\chi$ for a flat universe. In
terms of $H_0$ and $\Omega_0$, $K= (\Omega_0 -1)H_0^2$. In the
following we consider a small patch of the sky so we can use the
plane-parallel (or small-angle) approximation to replace spherical
harmonics by Fourier modes. The convergence $\kappa$ at a
direction $\gamma$ is given by a projection of the (3D) density
contrast along the line of sight with a weight function
$\omega(\chi)$ :
\begin{equation}
\kappa({\bf \gamma}) = \inc {d\chi} \omega(\chi)\delta(r(\chi),
{\bf \gamma}).
\end{equation}
If all the sources are at the same redshift, one can write the
weight function $\omega(\chi) = 3/4a c^{-2}H_0^2 \Omega_m r(\chi)
r(\chi_s - \chi)/ r( \chi_s)$,  where $\chi_s$ is the comoving
radial distance to the source. (This approximation is not crucial
and is easy to modify for a more accurate description.) Using a
Fourier decomposition of $\delta$ we can write
\begin{equation}
\kappa(\gamma) = \inc {d\chi} \omega(\chi) \int {d^3{\bf k} \over
{(2 \pi)}^3} \exp ( i \chi k_{\parallel} + r \theta k_{\perp} )
\delta_k,
\end{equation}
where we have used $k_{\parallel}$ and $k_{\perp}$ to denote the
components of wave vector $k$ parallel and perpendicular to the
line of sight. In small angle approximation however one assumes
that $k_{\perp}$ is much larger compared to $k_{\parallel}$,
$\theta$ denotes the angle between the line of sight direction
${\bf \gamma}$ and the wave vector ${\bf k}$.

\subsection{Cumulants}

Using the definitions we have introduced above we can compute the
projected two-point correlation function
\begin{equation}
\langle \kappa(\gamma_1) \kappa(\gamma_2) \rangle = \inc d
{\chi_1} {\omega^2(\chi_1) \over r^2(\chi_1)} \int {d^2 {\bf l}
\over (2 \pi)^2}~\exp ( i \theta l )~ {\rm P} { \left( {{\bf
l}\over r(\chi)} \right)}
\end{equation}
(Peebles 1980; Kaiser 1992; Kaiser 1998) where we defined ${\bf l}
= r(\chi){\bf k}_{\perp}$, a scaled wave-vector projected on the
surface of the sky. It is useful to consider the average of this
smoothed over an angle $\theta_0$,  which is often used to
reconstruct the matter power spectrum $P(k)$ (Jain, Selzak \&
White 1998). The convergence field smoothed by an angle $\theta_0$
is represented by
\begin{equation}
\kappa_{s}(\gamma) =  \int d^2 \gamma W_{\theta_0}(\gamma) \int_0^{\chi}
 d\chi \omega(\chi) \delta( r(\chi), \gamma),
\end{equation}
where $W_{\theta_0}$ is the ``top hat'' window function. We then
get
\begin{equation}
\langle \kappa^2_{s} \rangle = \inc d {\chi} {\omega^2(\chi) \over
r^2(\chi)} \int {d^2 {\bf l} \over (2 \pi)^2}~ P  \left( {l\over
r(\chi)} \right) W_2^2(l\theta_0),
\end{equation}
where $W_2(x)=2J_1(x)/x$ and $J_1(x)$ is a Bessel function; cf.
Bernardeau (1995). Similar calculations for the three- and
four-point correlation functions can be expressed in terms of
integrals of the matter multi-spectrum $B_p$, defined by
\begin{equation}
\langle \delta({\bf k}_1)\dots \delta({\bf k}_p)\rangle = B_p({\bf
k}_1, \dots {\bf k}_p),
\end{equation}
where the ${\bf k}_i$ are constrained to have $\sum {\bf k}_i =
0$. First define
\begin{equation}
I_N [r(\chi), \theta_0)]= \int {d^2 {\bf l_1} \over (2\pi)^2}
W_2(l_1 \theta_0) \dots \int {d^2 {\bf l_{ N-1 }} \over (2 \pi)^2}
W_2(l_{ N-1 }\theta_0) W_2(l_N\theta_0) B_N \left( {l_1\over
r(\chi)},\ldots, {l_N\over r(\chi)} \right)_{\sum {\bf l}_i = 0}.
\end{equation}
>From this we obtain
\begin{equation}
\langle \kappa^3_{s} \rangle = \inc d {\chi_1} {\omega^3(\chi)
\over r^6(\chi)} I_3 [r(\chi), \theta_0]
\end{equation}
and
\begin{eqnarray}
\langle \kappa^4_{s} \rangle = \inc d {\chi_1} {\omega^4(\chi)
\over r^8(\chi)} I_4 [r(\chi), \theta_0]
\end{eqnarray}
For the derivation of the corresponding hierarchical parameters,
\begin{equation}
 S_N = \frac{\langle \kappa^N_{s} \rangle}{\langle \kappa^{N-1}_{s} \rangle}
\end{equation}
 we need
\begin{eqnarray}
\langle \kappa^N_{s} \rangle = \inc d {\chi} {\omega^N(\chi) \over
r^{2(N-1)}(\chi)} I_N [r(\chi), \theta_0].
\end{eqnarray}
In these results we have assumed that the smoothing angle
$\theta_0$ is very small.

\subsection{Cumulant Correlators}

Cumulant correlators were introduced by Szapudi \& Szalay (1997)
as a normalized two-point moments of the  multi-point correlation
functions. Later the concept of two-point cumulant correlators was
generalized to multi-point cumulant correlators (Munshi, Melott \&
Coles 1999). The two-point cumulant correlators for galaxies have
already been measured in projected surveys such as APM by Szapudi
\& Szalay (1997) and they have been shown to be an important
diagnostic of hierarchical galaxy clustering models.

To calculate the two-point cumulant correlators  we adopt the
additional assumption that the separation between different angles
are small. This is the regime in which we expect the hierarchical
model presented in the next section to be most accurate. The
general result is:
\begin{eqnarray}
\langle \kappa_s^m(\gamma_1) \kappa_s^n(\gamma_2) \rangle =
\int_0^{\chi_s} { \omega^{n + m} (\chi) \over r^{2(n+m-1)}(\chi) }
d \chi \int \frac{d^2 {\bf  l}_1}{(2\pi)^2} \dots  \int
\frac{d^2{\bf l}_{n+m-1}}{(2\pi)^2} W_2(l_1 \theta_0)\dots W_2(
l_{n+m}\theta_0) \exp[i(l_1 + \dots + l_m)\theta] \nonumber \\
 B_{m+n} \left( {{\bf l}_1 \over r (\chi)},
 \dots, {{\bf l}_{m+n} \over r (\chi)} \right),
\end{eqnarray}
where again $\sum l_i = 0$, as it is in the following expressions.
Relevant low-order examples of this result are:
\begin{equation} \langle \kappa_s^2(\gamma_1) \kappa_s(\gamma_2)
\rangle =
 \int_0^{\chi_s} { \omega^3 (\chi) \over r^4(\chi) } d \chi \int
 \frac{d^2{\bf l}_1}{(2\pi)^2} \int  \frac{d^2{\bf l}_2}{(2\pi)^2}  W_2(l_1
 \theta_0) W_2(l_2 \theta_0) W_2( l_3\theta_0) \exp(il_2
 \theta_{12}) B_3 \left( {{\bf l}_1 \over r (\chi)},
 {{\bf l}_2 \over r (\chi)}, {{\bf l}_3 \over r (\chi)} \right);
\end{equation}
\begin{eqnarray}
\langle \kappa_s^3(\gamma_1) \kappa_s(\gamma_2) \rangle =
\int_0^{\chi_s} { \omega^3 (\chi) \over r^4(\chi) } d \chi \int
 \frac{d^2{\bf l}_1}{(2\pi)^2} \int  \frac{d^2{\bf l}_2}{(2\pi)^2}  \int  \frac{d^2{\bf
 l}_3}{(2\pi)^2} W_2(l_1 \theta_0) W_2(l_2 \theta_0) W_2( l_3\theta_0) \exp(il_3 \theta_{12})
  \nonumber\\
 B_4 \left( {{\bf l}_1 \over r (\chi)},
 {{\bf l}_2 \over r (\chi)}, {{\bf l}_3 \over r (\chi)}, {{\bf l}_4
\over r (\chi)} \right);
\end{eqnarray}
\begin{eqnarray}
\langle \kappa_s^2(\gamma_1) \kappa_s^2(\gamma_2) \rangle =
\int_0^{\chi_s} { \omega^3 (\chi) \over r^4(\chi) } d \chi \int
 \frac{d^2{\bf l}_1}{(2\pi)^2}  \int \frac{d^2{\bf l}_2}{(2\pi)^2}  \int  \frac{d^2{\bf
 l}_3}{(2\pi)^2}
W(l_1 \theta_0) W_2(l_2 \theta_0) W_2( l_3\theta_0)
\exp(i(l_1+l_2)\theta_{12})
 \\ \nonumber
 B_4 \left( {{\bf l}_1 \over r (\chi)},
 {{\bf l}_2 \over r (\chi)}, {{\bf l}_3 \over r (\chi)}, {{\bf l}_4
\over r (\chi)} \right).
\end{eqnarray}
The three-point cumulant correlators can be expressed in the
following form
\begin{eqnarray}
\langle \kappa_s^m(\gamma_1) \kappa_s^n(\gamma_2)
\kappa_s^r(\gamma_3) \rangle = \int_0^{\chi_s} {\omega^{n + m + r}
(\chi) \over r^{2(n+m+r-1)}(\chi) } d \chi \int  \frac{d^2{\bf
l}_1}{(2\pi)^2} \dots  \int  \frac{d^2{\bf l}_{n+m+r-1}}{(2\pi)^2}
W_2(l_1 \theta_0)\dots W_2( l_{n+m+r}\theta_0)
\\ \nonumber
 \exp[i(l_1 + \dots + l_m)\theta_{12}
+i(l_m + \dots + l_{m+n})\theta_{13}]
 B_{m+n+r} \left( {{\bf l}_1 \over r (\chi)},
 \dots, {{\bf l}_{m+n+r} \over r (\chi)} \right).
\end{eqnarray}

\section{Tree Model for Correlation Functions}

In deriving the expressions (9)-(17) we have not used any specific
form for the matter correlation hierarchy. To go further we have
to choose a particular form of hierarchical scaling. The most
general ``tree'' model can be written as
\begin{equation}
\xi_N( {\bf r_1}, \dots {\bf r_N} ) = \sum_{\alpha, \rm N-trees}
Q_{N,\alpha} \sum_{\rm labellings} \prod_{\rm edges}^{(N-1)}
\xi({\bf r_i}, {\bf r_j}).
\end{equation}
A similar hierarchy is developed in the quasi-linear regime in the
limit of vanishing variance (Bernardeau 1992), but the
hierarchical amplitudes $Q_{N, \alpha}$ are shape-dependent
functions in such a case. In the highly nonlinear regime of
interest here, there are some indications that these functions
become independent of shape (e.g. Scocciamarro et al. 1998). In
Fourier space, such a result means that the  multi-spectra can be
written
\begin{eqnarray}
B_2({\bf k}_1, {\bf k}_2, {\bf k}_3) & = & Q [ P({\bf k_1})P({\bf
k_2}) + P({\bf k_2})P({\bf k_3}) + P({\bf k_3})P({\bf k_1}) ]
\nonumber\\ B_3({\bf k}_1, {\bf k}_2, {\bf k}_3, {\bf k}_4) & = &
R_a [P({\bf k_1})P({\bf k_1 + k_2}) P({\bf k_1 + k_2 + k_3}) +
{\rm cyc. perm.}]\nonumber\\ & & + R_b [P({\bf k_1})P({\bf
k_2})P({\bf k_3}) + {\rm cyc. perm.}] \nonumber \\ B_4({\bf k}_1,
{\bf k}_2, {\bf k}_3, {\bf k}_4, {\bf k}_5) & = &  S_a [P({\bf
k_1})P({\bf k_1 + k_2}) P({\bf k_1 + k_2 + k_3})P({\bf
k_1+k_2+k_3+k_4})  + {\rm cyc. perm.} ]  \nonumber\\ & & + S_b [
P({\bf k_1})P({\bf k_2})P({\bf k_1+k_2+k_3})P({\bf
k_1+k_2+k_3+k_4})+ {\rm cyc. perm.}]\nonumber\\ & &  + S_c [P({\bf
k_1})P({\bf k_2})P({\bf k_3})P({\bf k_4}) + {\rm cyc. perm.}]
\end{eqnarray}
(remembering that $\sum {\bf k_i}=0$ in all these expressions.)

Particular hierarchical models differ in the way they fix the
various amplitudes for the different topologies ($Q$, $R_a$,
$R_b$, etc). For example, Bernardeau \& Schaeffer (1992)
considered the case where amplitudes can in general are
factorizable (see Munshi, Melott \& Coles for a detailed
description); see  Szapudi \& Szalay (1993) for an alternative
prescription. We will avoid any specific prescription because our
aim is to show that the statistics of $\kappa$ can allow one to
fix the hierarchical amplitudes empirically, as well as
constraining the power spectrum and background geometry. We shall
ignore smoothing corrections, for reasons discussed by Boschan,
Szapudi \& Szalay (1994). We first define
\begin{eqnarray}
\kappa_{\theta_0} & \equiv & \int  \frac{d^2\bf l}{(2\pi)^2} P
\left( {l \over r(\chi)} \right) W_2^2(l \theta_0), \nonumber\\
\kappa_{\theta_{12}} & \equiv & \int
 \frac{d^2\bf l}{(2\pi)^2} P \left( {l \over r(\chi)} \right)
W_2^2(l \theta_0) \exp ( l \theta_{12})\nonumber
\\ C_t[ F(\chi)] &  \equiv  & \int_0^{\chi_s} { \omega^{t} (\chi) \over
r^{2(t-1)}(\chi) } F(\chi) d \chi.
\end{eqnarray}
The expression for $C_t$ can be computed numerically for specific
models of background cosmology. $F(\chi)$ denotes the various
products of $\kappa_{\theta_0}$ and $\kappa_{\theta_{12}}$ which
appear in the following expressions containing $\chi$ dependence.
The values of hierarchical amplitudes (which in general are
insensitive to the background cosmology) can be computed from
numerical simulations or with analytic methods such as
hyper-extended perturbation theory (Scoccimarro et al. 1998;
 Scoccimarro \& Frieman 1998). The one-point cumulants can then be expressed as
\begin{eqnarray}
\langle \kappa^3(\gamma) \rangle & = & 3Q_3 C_3
[\kappa^2_{\theta_0}]  \nonumber \\ \langle \kappa^4(\gamma) \rangle
& = &  (12R_a + 4 R_b)C_4 [\kappa^3_{\theta_0}] \nonumber \\ \langle
\kappa^5(\gamma) \rangle & = &   (60S_a + 60 S_b+ 5S_c)C_5
[\kappa^4_{\theta_0}]
\end{eqnarray}
The first of these expressions was derived by Hui (1999). He
showed that this result agrees well with numerical ray-tracing
experiments by Jain, Seljak \&  White (1998). The expressions for
two-point cumulant correlators are
\begin{eqnarray}
\langle \kappa_s^2(\gamma_1) \kappa_s(\gamma_2) \rangle & = & 2Q_3
C_3 [\kappa_{\theta_0} \kappa_{\theta_{12}}] \nonumber \\ \langle
\kappa_s^3(\gamma_1) \kappa_s( \gamma_2) \rangle & = & (3R_a + 6
R_b)C_4 [\kappa_{\theta_0}^2 \kappa_{\theta_{12}}] \nonumber
\\  \langle \kappa_s^2(\gamma_1) \kappa_s^2(\gamma_2) \rangle & =
& 4 R_b C_4 [\kappa_{\theta_0}^2 \kappa_{\theta_{12}}] \nonumber \\
\langle \kappa_s^4(\gamma_1) \kappa_s(\gamma_2)\rangle & = &
(24S_a + 36S_b + 4 S_c)C_5 [\kappa_{\theta_0}^3
\kappa_{\theta_{12}}] \nonumber
\\ \langle \kappa_s^3(\gamma_1) \kappa_s^2(\gamma_2) \rangle & = &
 (12S_a + 6 S_b)C_5[\kappa_{\theta_0}^3 \kappa_{\theta_{12}}].
\end{eqnarray}
and the   three-point cumulant correlators are
\begin{eqnarray}
\langle \kappa_s^2(\gamma_1) \kappa_s(\gamma_2) \kappa_s(\gamma_3)
\rangle & = & C_5[ (4S_a \kappa_{\theta_{12}} \kappa_{\theta_{13}} +
4S_b \kappa_{\theta_{12}}\kappa_{\theta_{13}} + 4S_b
\kappa_{\theta_{13}}\kappa_{\theta_{23}}) \kappa_{\theta_0}^2]
\nonumber \\  \langle \kappa_s^3(\gamma_1) \kappa_s(\gamma_2)
\kappa_s(\gamma_3)  \rangle & = & C_5 [[(6S_a +
3S_b)(\kappa_{\theta_{12}}
\kappa_{\theta_{23}}+\kappa_{\theta_{13}} \kappa_{\theta_{23}})  +
(3S_c + 18 S_b + 6S_a)
\kappa_{\theta_{12}}\kappa_{\theta_{13}}] \kappa_{\theta_0}^2].
\end{eqnarray}
It is clear from these equations that at any order, measurements
of the appropriate cumulant correlators provide enough information
to determine the hierarchical coefficients of that order and
below. Weak lensing statistics can therefore tell us directly
about the form of hierarchical model appropriate for the
underlying density field, just as it can with the distribution of
galaxies (Munshi, Melott \& Coles 1999).

\section{Discussion}

Ongoing optical and radio experiments are intended to provide
imaging data for studies of weak lensing.  A large smoothing angle
is required to probe the quasi-linear regime where most existing
analytic calculations are valid. To be useful, such regions must
be at least reach 10 degrees on a side. Existing CCD cameras have
typical sizes ($0.25^{\circ}-0.5^{\circ}$). The initial lensing
surveys will therefore provide information about matter
fluctuations in the highly non-linear regime. This is the regime
in which the model we have presented here is expected to be valid.
Ongoing observational programmes in this regime allow the direct
determination of the hierarchical scaling properties of
cosmological density fluctuations. These can be combined with
similar measurements of galaxy clustering to constrain models of
the relationship between galaxies and mass. This is the key result
of this paper.

Other studies of small-scale  weak lensing have focussed on
ray-tracing experiments through N-body simulations. It will
definitely be very interesting to check our analytical results
against such numerical simulations.  This will also test different
approximations we have used in deriving our analytical results.
Results of this kind of analysis will be presented elsewhere
(Munshi, Jain \& White 1999a,b, in preparation).

Cumulant correlators of orders higher than those we have presented
here  can be written in a similar way to equations (21)-(23). We
have not presented them here, however, because sampling variance
plays an increasingly important role in these quantities at
increasing order, and we have not yet developed a method for
dealing with this in the context of weak lensing, although
estimation of finite-volume corrections has been done for galaxies
by several authors (Szapudi \& Colombi 1997; Munshi et al. 1999;
Munshi, Coles \& Melott 1999b; Hui \& Gaztanaga 1998). We also
note that the hierarchical ansatz we have exploited furnishes a
complete prescription of all the higher-order correlation
functions of the matter distribution. This allows one to estimate
the multi-point count probability distribution function for an
arbitrary number of correlated volume elements (Munshi, Coles \&
Melott 1999a,b; Bernardeau \& Schaeffer 1999). These results can
be used to compute the numbers of ``hot'' and ``cold'' spots in
statistics of $\kappa$ in the sky in much the same way as has been
done for fluctuations in the cosmic microwave background (e.g.
Coles \& Barrow 1987). The development of our work in these
directions will be presented elsewhere (Munshi \& Coles, 1999, in
preparion).

\section*{Acknowledgment}
DM was supported by a Humboldt Fellowship at the Max Planck
Institut fur Astrophysik when this work was performed.

\end{document}